\documentstyle[12pt,psfig]{article}

\def\EMPZ{$E-P_{\rm Z}$}
\newcommand{\PT}{{\mbox{$\not\hspace{-.55ex}{P}_{\rm t}$}}}
\def\Pt{$P_{\rm t}$}
\def\lsp{\chi_{1}^{0}}
\def\sel{\tilde e}
\def\squ{\tilde q}
\def\deps{\Delta\epsilon/\epsilon}

\def\3{\ss}
\begin{document}
\title {
 \hspace*{-5mm}\large\rm
       \LARGE  Search for Selectron and Squark Production \\
in $e^+p$ Collisions at HERA}

\author{ZEUS Collaboration}
\date{ }
\maketitle
\vspace{ 4cm}

\begin{abstract}

\noindent
We have searched for the production of a selectron and a squark
in $e^+p$ collisions at a center-of-mass energy of 300~GeV  using the 
ZEUS detector at HERA. The selectron and squark are sought in the direct decay
into the lightest neutralino in the framework of supersymmetric extensions to 
the Standard Model which conserve $R$-parity. No evidence for the production 
of supersymmetric particles has been found in a data sample corresponding to 
46.6~pb$^{-1}$ of integrated luminosity. We express upper limits on the 
product of the cross section times the decay branching ratios as excluded 
regions in the parameter space of the Minimal Supersymmetric Standard Model.
\end{abstract}

\vspace{-18cm}
{\noindent
  DESY-98-069 \newline
  June 1998}

\newpage

%
%
%
%
\newcommand{\address}{ }
\setcounter{page}{0}
\pagenumbering{Roman}                                                                              
                                                   %
\begin{center}                                                                                     
{                      \Large  The ZEUS Collaboration              }                               
\end{center}                                                                                       
  J.~Breitweg,                                                                                     
  M.~Derrick,                                                                                      
  D.~Krakauer,                                                                                     
  S.~Magill,                                                                                       
  D.~Mikunas,                                                                                      
  B.~Musgrave,                                                                                     
  J.~Repond,                                                                                       
  R.~Stanek,                                                                                       
  R.L.~Talaga,                                                                                     
  R.~Yoshida,                                                                                      
  H.~Zhang  \\                                                                                     
 {\it Argonne National Laboratory, Argonne, IL, USA}~$^{p}$
% ...
\newpage                                                                                           
$^{\    1}$ also at IROE Florence, Italy \\                                                        
% ...
$^{  20}$ now at ATLAS Collaboration, Univ. of Munich \\                                           
% ...
$^{  31}$ partially supported by the Foundation for German-Russian Collaboration                   
DFG-RFBR \\ \hspace*{3.5mm} (grant no. 436 RUS 113/248/3 and no. 436 RUS 113/248/2)\\              
% ...
                                                           %
                                                           %
\newpage   
                                                           %
                                                           %
\begin{tabular}[h]{rp{14cm}}                                                                       
$^{a}$ &  supported by the Natural Sciences and Engineering Research                               
          Council of Canada (NSERC)  \\                                                            
$^{b}$ &  supported by the FCAR of Qu\'ebec, Canada  \\                                            
$^{c}$ &  supported by the German Federal Ministry for Education and                               
          Science, Research and Technology (BMBF), under contract                                  
          numbers 057BN19P, 057FR19P, 057HH19P, 057HH29P \\                                        
$^{d}$ &  supported by the MINERVA Gesellschaft f\"ur Forschung GmbH,                              
          the German Israeli Foundation, the U.S.-Israel Binational                                
          Science Foundation, and by the Israel Ministry of Science \\                             
$^{e}$ &  supported by the German-Israeli Foundation, the Israel Science                           
          Foundation, the U.S.-Israel Binational Science Foundation, and by                        
          the Israel Ministry of Science \\                                                        
$^{f}$ &  supported by the Italian National Institute for Nuclear Physics                          
          (INFN) \\                                                                                
$^{g}$ &  supported by the Japanese Ministry of Education, Science and                             
          Culture (the Monbusho) and its grants for Scientific Research \\                         
$^{h}$ &  supported by the Korean Ministry of Education and Korea Science                          
          and Engineering Foundation  \\                                                           
$^{i}$ &  supported by the Netherlands Foundation for Research on                                  
          Matter (FOM) \\                                                                          
$^{j}$ &  supported by the Polish State Committee for Scientific                                   
          Research, grant No.~115/E-343/SPUB/P03/002/97, 2P03B10512,                               
          2P03B10612, 2P03B14212, 2P03B10412 \\                                                    
$^{k}$ &  supported by the Polish State Committee for Scientific                                   
          Research (grant No. 2P03B08614) and Foundation for                                       
          Polish-German Collaboration  \\                                                          
$^{l}$ &  partially supported by the German Federal Ministry for                                   
          Education and Science, Research and Technology (BMBF)  \\                                
$^{m}$ &  supported by the Fund for Fundamental Research of Russian Ministry                       
          for Science and Edu\-cation and by the German Federal Ministry for                       
          Education and Science, Research and Technology (BMBF) \\                                 
$^{n}$ &  supported by the Spanish Ministry of Education                                           
          and Science through funds provided by CICYT \\                                           
$^{o}$ &  supported by the Particle Physics and                                                    
          Astronomy Research Council \\                                                            
$^{p}$ &  supported by the US Department of Energy \\                                              
$^{q}$ &  supported by the US National Science Foundation \\                                       
\end{tabular}                                                                                      
                                                           %
                                                           %
\newpage
\pagestyle{plain}                   
\thispagestyle{empty}
\normalsize
\pagenumbering{arabic}
\setcounter{page}{1}

\section{Introduction}
\label{sec:intro}

Supersymmetry (SUSY) theories relate bosons and fermions by associating to 
each fermion a bosonic partner and vice-versa. Among the appealing
consequences of this symmetry is the cancellation of quadratic
divergences occurring in the scalar Higgs sector of the Standard Model
(SM) and models beyond the SM \cite{newsusy0}.
There is, on the other hand, no experimental evidence for SUSY.
Since supersymmetric particles are not observed at the masses of their 
standard partners, SUSY would need to be broken.

In the supersymmetric extension of the SM known as the Minimal Supersymmetric 
Standard Model (MSSM),
the supersymmetry-breaking terms are added by hand, generating a 
model with many free parameters (for a review and the original references
see \cite{mssm}).
In this model  
the breaking of ${\rm SU}(2)\times {\rm U}(1)_{\rm Y}$ is generated through 
the vacuum expectation values
$v_1$ and $v_2$ of two Higgs doublets, which give masses to the down-type
quarks and charged leptons ($v_1$) and to the up-type quarks ($v_2$).
Selectrons ($\tilde e_{\rm L}$, $\tilde e_{\rm R}$) 
and squarks ($\tilde q_{\rm L}^{\rm f}$, $\tilde q_{\rm R}^{\rm f}$) are
the scalar partners of the left- and right-handed electrons and
the quarks of flavor ${\rm f}$.
The supersymmetric partners of the 
gauge bosons and the Higgs particles, known as gauginos and higgsinos, 
mix together
giving rise to four neutral mass eigenstates $\chi^{0}_{i}$ (neutralinos) and
two charged mass eigenstates $\chi^{\pm}_j$ (charginos).
The masses of the neutralinos depend on the supersymmetry-breaking 
parameters $M_1$ and
$M_2$ (for the U(1) and SU(2) gauginos), on the higgsino mass parameter $\mu$
and on the ratio of the two Higgs vacuum expectation values 
$\mbox{tan}\beta \equiv v_2/v_1$. 
There is a new multiplicative quantum number, $R$-parity ($R_P$), 
which takes the
values +1 for the ordinary particles and -1 for the supersymmetric
particles.
In models where $R_P$ is conserved, the supersymmetric particles are
produced only in pairs and the lightest supersymmetric particle 
(LSP) is stable. 
In these models, the production of a slepton and a squark is 
the lowest order process in which supersymmetric particles could be produced
at HERA \cite{sqse1,sqse2,sqse3}. 
Other processes, 
such as the production of a slepton and a gaugino \cite{z} have a smaller
cross section.

\begin{figure}
\centerline{
\hbox{
\psfig{file=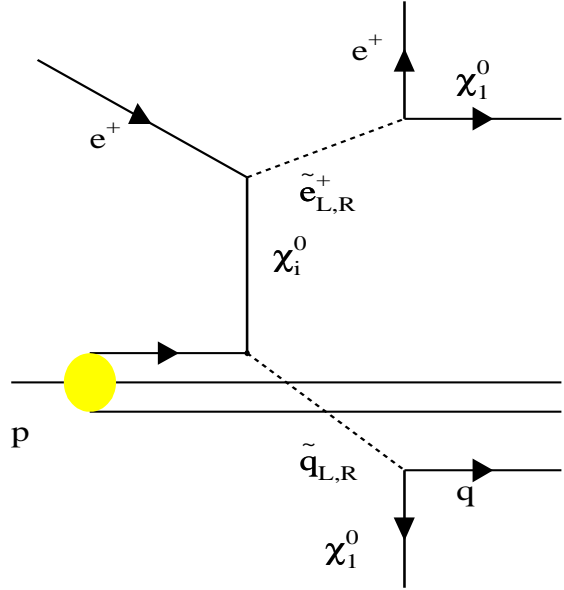,height=13cm,width=13 cm}
}}
\caption{\label{f0} Diagram for the selectron and squark 
production process.}
\end{figure}

We report here on the search for the $R_P$-conserving production 
and decay of a selectron~\footnote{Throughout this paper we will call 
``selectron'' the scalar partner
of the positron.} and a squark,
$ e^+ p \rightarrow \tilde e^+_{\rm a} \tilde q_{\rm b}^{\rm f}X $  
(${\rm a=L,R}$, ${\rm b=L,R}$). 
This process is mediated by the $t$-channel exchange of a neutralino, as shown 
in figure~\ref{f0}.
As the production of high-mass particles in the final state 
involves high Bjorken-$x$ valence quarks from the proton, the process
is mostly sensitive to the up quark.
This measurement was performed with the
ZEUS detector using an integrated luminosity ${\cal L}\,=\,46.6\,\rm{pb}^{-1}$.
The H1 collaboration has published results of 
a search similar to the one presented in this paper, using
an integrated luminosity ${\cal L}\simeq 6.4 \,{\rm pb}^{-1}$ \cite{h1}.

The selectron (squark) can decay directly to the lightest neutralino,
$\chi^{0}_{1}$, and
a positron (quark):
$\tilde e \rightarrow e \chi^{0}_{1}$  
($\tilde q \rightarrow q \chi^{0}_{1}$). 
Under the assumption that the lightest neutralino is the LSP and that $R_P$
is conserved, one can conclude that the neutralino
 will escape detection.
In this case, the signature
for the production of a selectron and a squark is one positron from
$\sel$ decay, a high \Pt ~hadronic system from $\squ$ decay and missing
momentum from the two neutralinos \cite{cashmore}.
The search for this process is used to set limits on the masses of the 
selectron and of the squark
for a wide range of values of the MSSM parameters. 
Limits are derived for all squarks assuming that they
are degenerate in mass. We also set limits separately for the $\tilde u$ 
squark. 
These limits can be considered complementary to the strong limits
on the squark mass from $p\bar{p}$ collider experiments \cite{tevatron}
that are valid within
models with additional constraints compared to the MSSM. 
The present search leads to limits comparable to those so far obtained 
from $e^+e^-$ experiments \cite{lep,stop}
for particular combinations of the $\tilde e$ and $\tilde q$ masses.

\section{Experimental Setup}
\label{sec:setup}

The $46.6\,{\rm pb}^{-1}$ of data used for this analysis 
were collected with the ZEUS detector at the HERA
collider during the years 1994 to 1997. The collider operated 
at a center-of-mass energy of 300~GeV with
positrons of energy $E_e = 27.5\,\mbox{GeV}$ and protons of energy 
$E_p = 820\,\mbox{GeV}$.

The ZEUS detector is described in detail elsewhere
\cite{zeusdet}. The main subdetectors used in the present analysis are the
central tracking detector (CTD) \cite{ctd} 
positioned in a 1.43 T solenoidal magnetic 
field and the compensating uranium-scintillator sampling calorimeter (CAL) 
\cite{cal},
subdivided in forward (FCAL), barrel (BCAL) and rear (RCAL) sections.
Under test beam conditions the CAL energy resolution is 
$18\%/\sqrt{E \mbox{(GeV)}}$ for
electrons and $35\%/\sqrt{E\mbox{(GeV)}}$ for hadrons. 
A three-level trigger was used to select events online.
The trigger criteria applied relied primarily on the energies deposited in the
calorimeter. The trigger decision was based on electromagnetic
energy, total transverse energy and missing transverse momentum.
Timing cuts were used to reject beam-gas interactions and cosmic rays.

The luminosity was measured to a precision of 1.5\% from the rate of
energetic brems\-strah\-lung photons produced in the process $e^+\,p\rightarrow
e\,p\,\gamma$ \cite{lumi}.

\section{Production Model}
\label{sec:model}

The amplitude of the process depicted in figure \ref{f0}
is given by the sum of four exchange graphs, one 
for each MSSM neutralino. The cross section depends on the MSSM parameters
$M_1$, $M_2$, $\mu$, $\tan \beta$, 
and on the masses of the produced particles, 
$ \sigma_{ep \rightarrow \tilde e_{\rm a} \tilde q^{\rm f}_{\rm b} X}=
 \sigma_{\rm ab}^{\rm f}(M_{1},M_{2},\mu,\tan\beta,m_{\tilde e_{\rm a}},
m_{\tilde q_{\rm b}^{\rm f}}).$
The cross section depends to a very good approximation on the sum 
$m_{\sel}+m_{\squ}$ of the scalar particle masses.
The branching ratios, $B$, for the decays $\sel\rightarrow e\lsp$ and 
$\squ\rightarrow q\lsp$ depend on the same MSSM parameters.
Squarks are not allowed to decay to gluinos ($\tilde g$) as we assume 
$m_{\tilde g}>m_{\squ}$.

For $|\mu|<M_1$ and $M_2$, the $\lsp$ is higgsino-like and the cross sections 
are small, due to the small coupling constants. 
In the region $|\mu|>M_1$ and $M_2$, the
$\lsp$ is gaugino-like and good limits on the sparticle masses are
possible with the exception of particular combinations of $M_1$ and $M_2$ 
(for $\mu > 0$) when the LSP becomes a chargino. 
In the asymptotic region $|\mu|\gg M_2 > M_1$, the mass of the lightest
neutralino is
$m_{\lsp}\sim M_1$ and the masses of the next-to-lightest neutralino 
and of the charginos are $m_{\chi^0_2}\sim m_{\chi^+_1}\sim M_2$; therefore
for $M_2>m_{\sel}$ and $m_{\squ}$, the scalar particles can only
decay to the LSP and $B=1$, while $B\approx 30\%$ 
when $M_2 < m_{\sel}$ and $m_{\squ}$.
In this region the cross section depends weakly on $\tan\beta$. 
We take $|\mu|=500\,{\rm GeV}$ as an example of large $\mu$ values. 

To reduce the number of free parameters, the $\sel_{\rm L}$ and $\sel_{\rm R}$ 
are assumed to have the same mass $m_{\tilde e}$, and all the squarks 
(except $\tilde t$ whose contribution can be neglected) to have the same mass 
$m_{\tilde q}$. 
Alternatively we consider the case that $\tilde u$ is the only squark 
contributing to the process.
We assume no mixing between  $\sel_{\rm L}$ ($\squ_{\rm L}^{\rm f}$) and  
$\sel_{\rm R}$ ($\squ_{\rm R}^{\rm f}$).
In Grand Unified Theories (GUT) the gaugino masses unify at the GUT scale
which leads, at the electroweak scale, to the relation 
$M_1\,=\,{5\over 3}\tan^2\theta_W\,M_2 \,\simeq\,{1\over 2} M_2$, 
which is adopted here to fix the branching ratios and 
compare with other experiments.

\section{Monte Carlo Simulation}
\label{sec:monte}

Monte Carlo simulations are used to determine the efficiency for 
selecting the signal, and also to estimate the rate of the SM background.
All the signal and background events were passed through the simulation 
of the detector response based on GEANT \cite{geant}, incorporating the 
effects of the trigger.
They were subsequently processed with the same reconstruction and
analysis programs used for the data.

\subsection{Signal}
The calculation of the leading-order
 cross sections and the simulation of the signal
have been performed with a Monte Carlo (MC) generator 
(MSSM, see \cite{mssmgen}) based on the calculations of \cite{sqse3}.
In the calculation of the total cross section, the proton parton density
is modeled using the GRV-LO \cite{GRV} set.
The change of the structure functions to CTEQ-LO \cite{cteq} and GRV-HO
produced variations always less than 4\%.
The effect of the initial state QED radiation from the incoming lepton was
included using the Weizs\"acker-Williams approximation.
To generate the signal MC samples
the MSSM MC is interfaced to ARIADNE \cite{ariadne}, which implements the
Color Dipole Model (CDM) to simulate the QCD radiation, 
and JETSET \cite{jetset}, for the fragmentation of the final state partons.
The signal samples
were generated for 118 different combinations of the values for
the masses $m_{\tilde e}$,
$m_{\tilde q}$ and $m_{\chi^0_1}$, by assuming the exchanged neutralino
to be a pure photino ($\tilde \gamma$), since the event kinematics and the
distributions of the decay products are insensitive to the neutralino mixing
parameters. 
The $\tilde e$ and $\tilde q$ were then forced to decay
directly into $\lsp$.

\subsection{Background}
To evaluate the SM background
we considered the following reactions which can mimic the signature for a 
selectron-squark event: 
\begin{itemize}
\item neutral current (NC) deep inelastic scattering (DIS) events with 
missing momentum due to the presence of neutrinos, muons 
or not-completely-contained hadronic energy;
\item charged current (CC) DIS events with a true or fake positron in the 
final state;
\item $W$ production processes which can mimic the signal via the leptonic 
decays;
\item lepton pair production from Bethe-Heitler interactions
($\gamma\gamma\rightarrow e^+e^-,\mu^+\mu^-,\tau^+\tau^-$);
\item photoproduction with high transverse energy and heavy flavor production, 
with true or fake positrons in the final state.
\end{itemize}

Backgrounds from NC and CC DIS events were simulated using HERACLES
\cite{heracles}, which includes first order
electroweak radiative corrections. The hadronic final states 
were simulated using LEPTO \cite{newlepto} and
JETSET interfaced with HERACLES using DJANGO \cite{django}. QCD radiation
was simulated using the CDM as implemented in ARIADNE
and, as a cross check, with the exact first order matrix
elements followed by parton showering (MEPS option).
The production of $W^{\pm}$ was simulated using EPVEC \cite{epvec}.
The production of lepton pairs in both elastic and inelastic Bethe-Heitler
interactions is simulated using LPAIR \cite{lpair}. 
All these samples correspond to luminosities at least five
times larger than that of the data.
The photoproduction background was simulated using events with
high transverse energy produced in direct- and resolved-photon 
interactions, which were generated using HERWIG \cite{herwig}.
Samples of heavy flavors ($\bar{c} c$,
$\bar{b} b$) were also produced via the boson gluon fusion mechanism, as 
implemented in AROMA \cite{aroma}. These last two samples, which represent
a small background, correspond to a
luminosity comparable to that of the data.

\section{Event Selection}
\label{sec:kinema}

The event selection for the SUSY particles search requires the following
global event quantities calculated from the reconstructed vertex 
and the calorimeter measurements:
\begin{eqnarray}
E-P_{{\rm Z}} &=&{\ \sum_{\rm i}}\, \left( E_{{\rm i}}-P_{{\rm Z,i}}\right) 
\nonumber \\
\not{\hspace{-.55ex}}{P}_{{\rm t}} &=&\sqrt{ \left( {\textstyle \sum_{\rm i}}\, P_{
{\rm X,i}}\right) ^2 + \left( {\textstyle \sum_{\rm i}}\, P_{{\rm Y,i}} \right) 
^2} \nonumber
\end{eqnarray}
where $E_{\rm i}$ is the energy measured in the i$^{th}$ calorimeter cell and
${\vec P}_{\rm i}\equiv E_{\rm i}\,{\vec n}_{\rm i}$, with ${\vec n}_{\rm i}$ the
unit vector from the reconstructed vertex to the cell 
center\footnote{The ZEUS coordinate system is a right handed
system with the Z axis pointing in the proton direction. 
The pseudorapidity is defined as
$\eta=-\ln(\tan{\theta\over 2})$, and the polar angle $\theta$ is measured
with respect to the Z direction.}.
The sums run over all calorimeter cells. 
The missing transverse momentum 
{\mbox{$\not\hspace{-.55ex}{P}_{\rm t}$}}\
and the longitudinal momentum variable $E-P_{{\rm Z}}$ equal, respectively,
zero and twice the incident positron beam energy ($2E_e=55$~GeV) 
for fully contained events. The signal events are expected to have large
{\mbox{$\not\hspace{-.55ex}{P}_{\rm t}$}}\ and low $E-P_{{\rm Z}}$ due to
the momentum carried away by the escaping neutralinos.
The calculation of the hadronic transverse momentum ($P^{\rm h}_{\rm t}$) 
excludes cells in the
calorimeter sum which belong to a cluster which is identified as the candidate 
positron from $\tilde{e}$ decay.

Events are selected by requiring an identified positron
with transverse momentum $P_{\rm t}^{\rm e} > 4$ GeV and a hadronic 
system with $P_{\rm t}^{\rm h}> 4$ GeV.

Positron
candidates were identified using the pattern of energy deposits in CAL
\cite{sinistra}, which provided the measurement of its energy $E_e$ and
angles $\theta_e$, $\phi_e$.
Additional requirements were applied:

\begin{itemize}

\item {\it isolation}: the total transverse energy not associated with the 
positron candidate within a
cone of radius R=0.8 units in the ($\eta,\phi$) plane around its 
direction must be less
than $3\,\rm{GeV}$ and the sum of the momenta of the tracks pointing in a cone
of radius R=0.8 units in $\eta$,$\phi$ around the positron, excluding the 
highest momentum track, must be less than 2 GeV. These cuts remove fake
positrons from CC and photoproduction that are close to the hadronic jet;

\item {\it track matching}: if the candidate positron is within a region 
where the tracking efficiency is high, namely 
$\theta_e\,>\,0.35\,\rm{radians}$, a match with a track is required such that

 \begin{itemize}

  \item the distance of closest approach between the electromagnetic (EM) 
        cluster and the track extrapolated from the CTD be less than 6 cm,

  \item the ratio of the track to CAL transverse momentum, 
         $P^{\rm track}_{\rm t}/P^{\rm e}_{\rm t}\,>\,0.2$,

  \item the track has a positive charge or $P^{\rm track}_{\rm t}\,>\,45\,\rm{GeV}$;

 \end{itemize}

if $\theta_e \,<\,0.35$, the requirement on the candidate positron
transverse momentum is raised to $P^{\rm e}_{\rm t}\,>\,10\,\rm{GeV}$;

\item {\it fiducial cut}: polar angle $\theta_e\,<\,2$; 
this cut eliminates most of the low $Q^2$ NC DIS background.
\end{itemize}

To ensure that the selected events come from $e^+p$ interactions 
we apply standard cosmic ray and beam gas rejection cuts and require
a reconstructed vertex within 50 cm of the nominal interaction point along
the Z axis.

\begin{figure}
\centerline{
\hbox{
\psfig{file=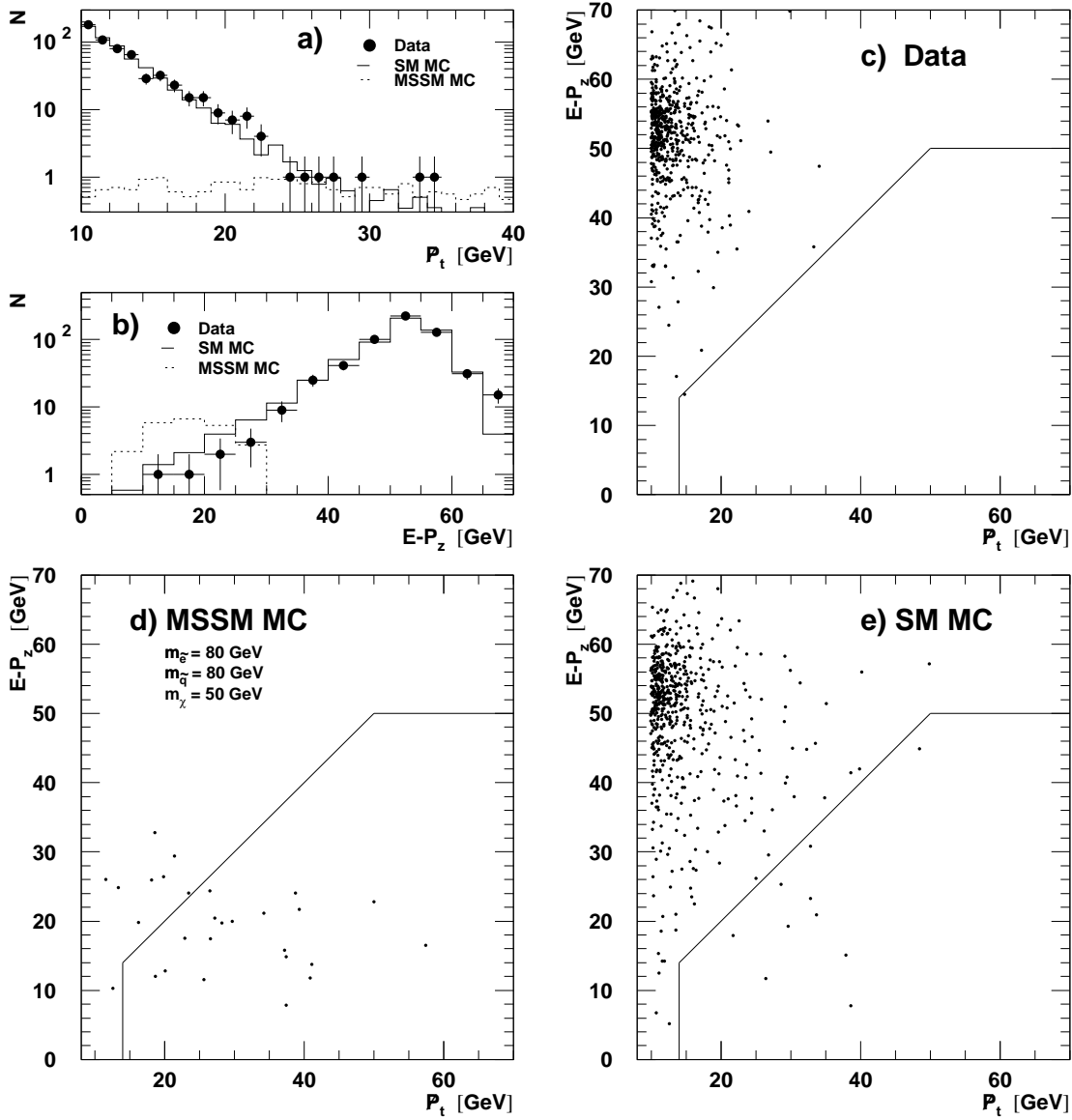,height=17cm}
}}
\caption{\label{f1} Distributions of events with a selected positron
and \PT$>10 \,{\rm GeV}$, in a) \PT ~and  b) \EMPZ ~for data
(points), SM background (full line) and a signal example
 ($m_{\tilde e}=80$ GeV,
$m_{\tilde q}=80$ GeV, $m_{\chi_{1}^{0}}=50$ GeV) (dashed line).
The distributions of events in the  \EMPZ~versus~\PT ~plane are shown in c) 
for the data, in d) for the MSSM example and in e)
for the SM background, where the MC samples
are normalized to 5 times the luminosity of the data.}
\end{figure}

In figures \ref{f1}~a) and b) the \PT ~and \EMPZ ~distributions
of the data, after a cut \PT$\,>\,10\,\rm{GeV}$, (full points) 
are compared with the total background expectation (full line) and with
a Monte Carlo
signal sample corresponding to the masses $m_{\tilde e}\,=\,80\,\rm{GeV}$,
$m_{\tilde q}\,=\,80\,\rm{GeV}$ and $m_{\chi^0_1}\,=\,50\,\rm{GeV}$ 
(dotted line).
Good agreement is observed between the distributions of the
data and the background Monte Carlo.

To determine the values of the final cuts, an optimization procedure was 
applied, which maximized the ratio ${\epsilon/{\rm U}}$ for simulated events, 
where $\epsilon$ is the efficiency and
${\rm U}$ is the 95\% confidence level (CL) upper limit on the signal when
the number of events found equals the expected background. Events were
accepted if they satisfied the following requirements:

\begin{itemize}
\item \EMPZ $\,<\,50\,\mbox{GeV}$;
\item \PT $\,>\,14\,\mbox{GeV}$;
\item (\EMPZ)/\PT $\,<\,1$.
\end{itemize} 

The acceptance for the $e^+\,q\,\chi^0_1\,\chi^0_1$ final state is 
close to zero for small mass differences 
$\Delta m \,=\,\mbox{min}(m_{\sel}-m_{\lsp},m_{\squ}-m_{\lsp})$
and reaches a plateau for $\Delta m> 10\,\mbox{GeV}$.
The level of the plateau 
increases from 25\% at 
$m_{\tilde e}=m_{\tilde q}=40\,\rm{GeV}$ to about 50\% 
at $m_{\tilde e}$ or $m_{\tilde q}=120\,\rm{GeV}$.
The acceptance was modeled with a
two parameter fit as a function of $m_{\chi^0_1}$ for 
fixed $m_{\tilde e}$ and $m_{\tilde q}$. Subsequently a bilinear
interpolation was used to obtain the parameters at any $m_{\tilde e}$ and 
$m_{\tilde q}$.

In figures \ref{f1}~c),~d) and e),
the data, the signal sample and the SM background are shown in the 
\PT~,~\EMPZ ~plane, where the final cuts are also drawn.
The expected SM background
is $1.99^{+0.57}_{-0.84}$ events, as shown in table \ref{t1}. 
The systematic uncertainties quoted for the background are evaluated as
described in section~\ref{sec:system}. 
No photoproduction or heavy flavor event survived the selection. 
An alternative method was used to confirm that the photoproduction
background was not
underestimated due to the limited MC statistics. It was found that the
rate of photoproduction events passing all of the kinematic cuts 
(except positron identification) was only a small fraction of the CC
background rate. Assuming that the fake positron fraction was the same
for CC and photoproduction events, the resulting photoproduction rate
was negligible.

One event survived the selection criteria and was identified as
containing a high-$Q^2$ positron with associated \PT~ in the calorimeter
due to two muons in the final state.

\section{Systematic Uncertainties}
\label{sec:system}

The main source of systematic uncertainty on the efficiency $\epsilon$
comes from the interpolation between the generated Monte Carlo mass points
($\deps = 6\%$). Other contributions from detector effects that affect 
both the
acceptance and the background are the uncertainties on
the positron identification efficiency ($\deps = 1\%$), on
the calorimeter energy scales ($\deps = 2\%$), on the 
efficiency of the matching between a track and an EM cluster 
($\deps = 3.5\%$),
on the isolation requirements ($\deps = 3\%$) and on the
vertex distribution ($\deps = 1\%$). Another common source is the 1.5\%
uncertainty in the total integrated luminosity of the 1994-1997 $e^+p$
data sample. We quote an overall systematic uncertainty $\deps = 8.5\%$.

In addition to the experimental sources, effects due to 
the different treatment of the QCD radiation in the final state for the
CC and NC background were evaluated using Monte Carlo samples generated
with the MEPS option.
The variation of the CC background was found to be below the statistical 
uncertainty. No NC MEPS event survived the final selection. This difference
with respect to the ARIADNE NC background is included as a systematic 
uncertainty.
The theoretical uncertainties on the cross sections of
the background processes are also included in table \ref{t1}.
The quoted errors are due to uncertainties on: the proton parton distribution
functions (pdf) for the NC and CC backgrounds ($5-6\%$), the missing
higher orders and proton and photon pdf for W production ($35\%$) and 
the uncertainties mainly due to inelastic contributions on the  lepton pair 
production ($10\%$).
The total error on the background ($\sim \pm 35\%)$ is taken to be the sum in 
quadrature of the statistical and systematic errors.
For the estimated background and one
observed event this $\sim 35\%$ variation of the background normalization
results in a 4\% variation of the
95\% CL upper limit on the number of signal events.

\section{Results}
\label{sec:results}

\begin{figure}[hp]
 \centerline{
    \hbox{
      \psfig{figure=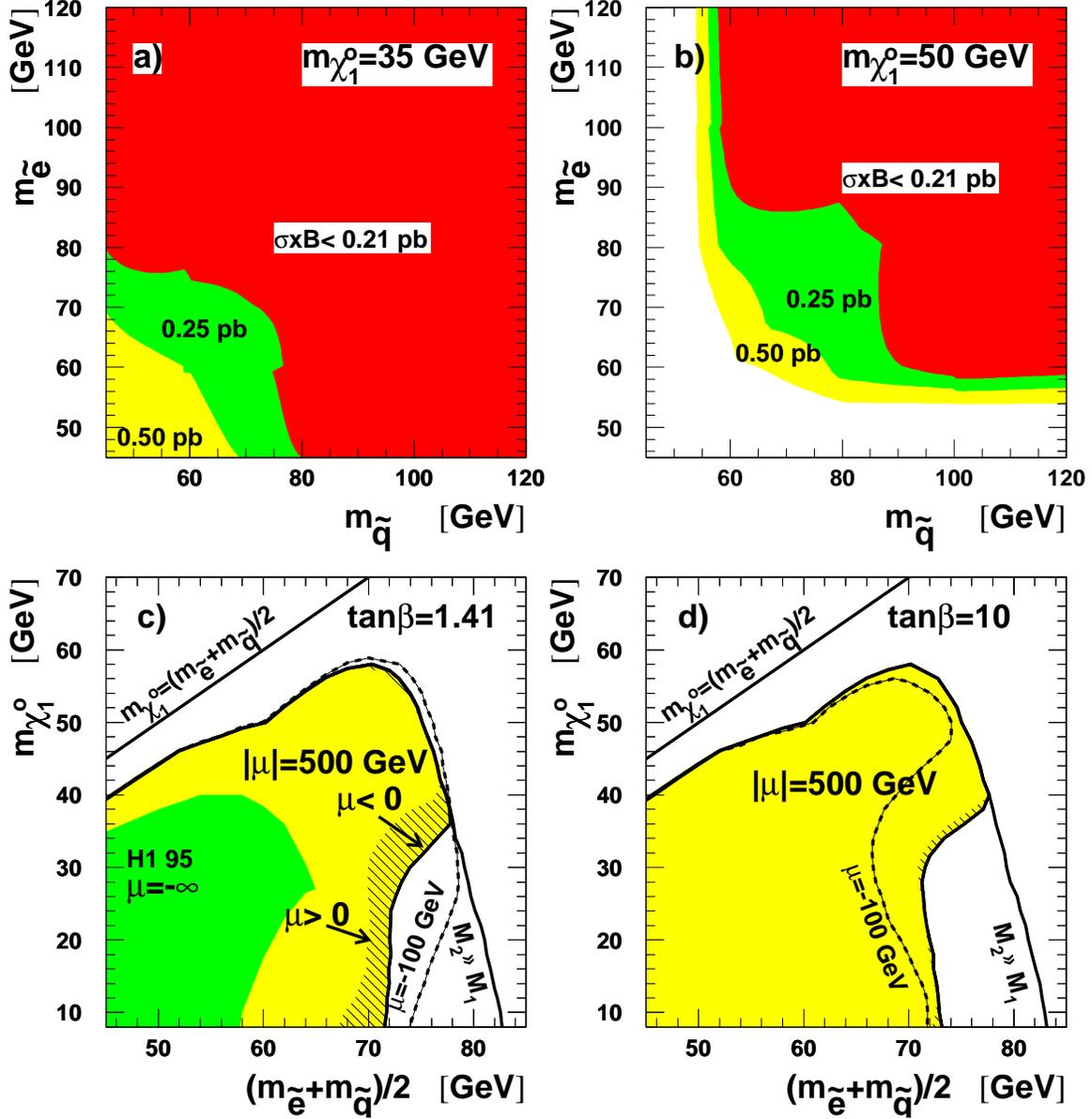,height=17cm}
                    }
                                 }
                                 
         \caption{\label{f2} a) and b) 95\% CL upper limits on 
         $\sigma \times B$
         in the plane defined by $m_{\squ}$ and $m_{\sel}$ for
         two values of the mass of the lightest neutralino:
         $m_{\lsp}=35$ GeV in a) and $m_{\lsp}=50$ GeV in b).
         c) and d) show the regions excluded at the 95 \% CL in
         the plane defined by  $(m_{\sel}+m_{\squ})/2$ and $m_{\lsp}$, 
         where limits are evaluated along $m_{\sel}=m_{\squ}$ 
         for $\tan \beta = 1.41$ (c)) and for $\tan \beta = 10$ (d)).
         The grey area is for $|\mu|=500$ GeV. For $\mu<0$
         the excluded region covers also the hatched grey area. 
         The dashed line is for $\mu=-100$ GeV.
         These limits are obtained for $M_{2}=2M_{1}$. If this relation 
         is modified to 
         $M_{2}\gg M_{1}$ the excluded area for $|\mu|=500$  
         reaches the full line. 
         The kinematic limit is indicated by the straight line in the
         upper-left corner.
         Previous H1 limits from \cite{h1} are also shown in c).
                   }
         \end{figure}

The event found in the data sample is compatible with the expected SM 
background. Therefore we derive upper limits  on the $\tilde
e$,$\tilde q$ production cross section times the branching ratios
($\sigma \times B$) for the decay to the lightest neutralino $\chi^0_1$.
The 95\% CL upper limit on the signal, including the 
systematic uncertainty on the acceptance and the uncertainty in the
background as described in \cite{estar},
is $N_{\rm UL}=3.9$ events.
The 95\% CL upper limit on  
$\sigma \times B$ is obtained as a function of the selectron,
squark and neutralino masses through
$\sigma \times B\,<\,{N_{\rm UL} / \epsilon {\cal L}}$, 
where $\epsilon\,=\,\epsilon(m_{\sel},m_{\squ},m_{\lsp})$ is the 
parametrized efficiency and ${\cal L}$ is the 
integrated luminosity. Figures \ref{f2} a) and b) show the upper limit
on  $\sigma \times B$ in the plane defined by $m_{\sel}$ and $m_{\squ}$
for two values of $m_{\lsp}$ (35 and 50 GeV). 
At high $m_{\sel}$ and $m_{\squ}$ and large $\Delta m$ we exclude
$\sigma \times B > 0.21 \,{\rm pb}$.
For $m_{\lsp} \le 50$ GeV, $\Delta m \,>\,10\,\mbox{GeV}$ and
$(m_{\sel}+m_{\squ})/2 > 70$ GeV we exclude
$\sigma \times B\,>\,0.25\,{\rm pb}$. For any mass combination with 
$\Delta m > 10 \,{\rm GeV}$ we exclude $\sigma \times B\,>\,0.5\,{\rm pb}$.

From the comparison of the upper limit on $\sigma \times B$ with the
theoretical value from the model calculations we
derive exclusion areas in the parameter space of the MSSM. 
The theoretical cross sections were evaluated
by taking into account the complete neutralino mixing and the 
branching ratios for the decays of $\tilde e$ and $\tilde q$ to $\chi^0_1$,
computed at the tree level.
The cross section times branching ratios is computed from:
\begin{eqnarray*} 
 (\sigma \times B)_{\rm th} \;=\,
\sum_{\rm a,b=L,R}\sum_{\rm f=1}^{5}\, 
 \{ \,\!\!\!\!&\sigma_{e^+p \rightarrow \sel_{\rm a}\squ_{\rm b}^{\rm f}}&\!\!\!\!
{ B}(\sel_{\rm a}^+\rightarrow e^+\lsp)\,{ B}(\squ_{\rm b}^{\rm f}\rightarrow 
q^{\rm f}\lsp)+ \nonumber \\ 
 \!\!\!\!&\sigma_{e^+p \rightarrow \sel_{\rm a}\bar{\squ}_{\rm b}^{\rm f}}&
 \!\!\!\!{ B}(\sel_{\rm a}^+\rightarrow e^+\lsp)\,
 { B} (\bar{\squ}_{\rm b}^{\rm f} \rightarrow \bar{q}^{\rm f}\lsp)\,\}.
\end{eqnarray*}
In figures~\ref{f2} c) and d) we show the excluded regions 
in the plane defined by the lightest neutralino and by 
$(m_{\sel}+m_{\squ})/2$. 
The cross section for the $\sel$ and $\squ$ production depends
mainly on the sum of the two masses; we give the excluded region for 
$m_{\sel}=m_{\squ}$ since the efficiency and the branching ratios depend 
separately on the two masses.
The limits are shown for $\mu = \pm 500$ GeV as examples of large values 
of $\mu$. The limits for $\mu=+500$ GeV and $\mu=-500$ GeV are similar and
differ by the hatched region in the figures. Limits are also shown 
for the intermediate value $\mu=-100\,{\rm GeV}$. 
For large $|\mu|$ the excluded
region reaches $(m_{\tilde e}+m_{\tilde q})/2\,=\,77\,\mbox{GeV}$ for a 40 GeV
neutralino. This limit worsens
at lower neutralino masses, because new decay channels to charginos and next
to lightest neutralino open and compete with the direct decay to $\chi^0_1$.
In the limit $M_2 \gg M_1$ the charginos and the next to lightest
neutralino masses increase leaving only the direct decay channel to 
$\lsp$ open.
In this case the excluded region reaches $(m_{\sel}+m_{\squ})/2=84$ GeV 
for a massless neutralino.
The excluded region is limited by the small cross section of the process
at large $(m_{\sel}+m_{\squ})/2$, while for large neutralino masses it is 
limited by the efficiency that falls to zero as $\Delta m \rightarrow 0$.
A large variation in $\tan\beta$ results only in slight changes.
The limits are shown in c) for $\tan\beta=1.41$ and in d) for $\tan\beta=10$ 
as examples of a small and a large value of $\tan\beta$. 
The previous limits obtained by H1 are also shown in figure~\ref{f2} c).

\begin{figure}[hp]
 \centerline{
    \hbox{
          \psfig{figure=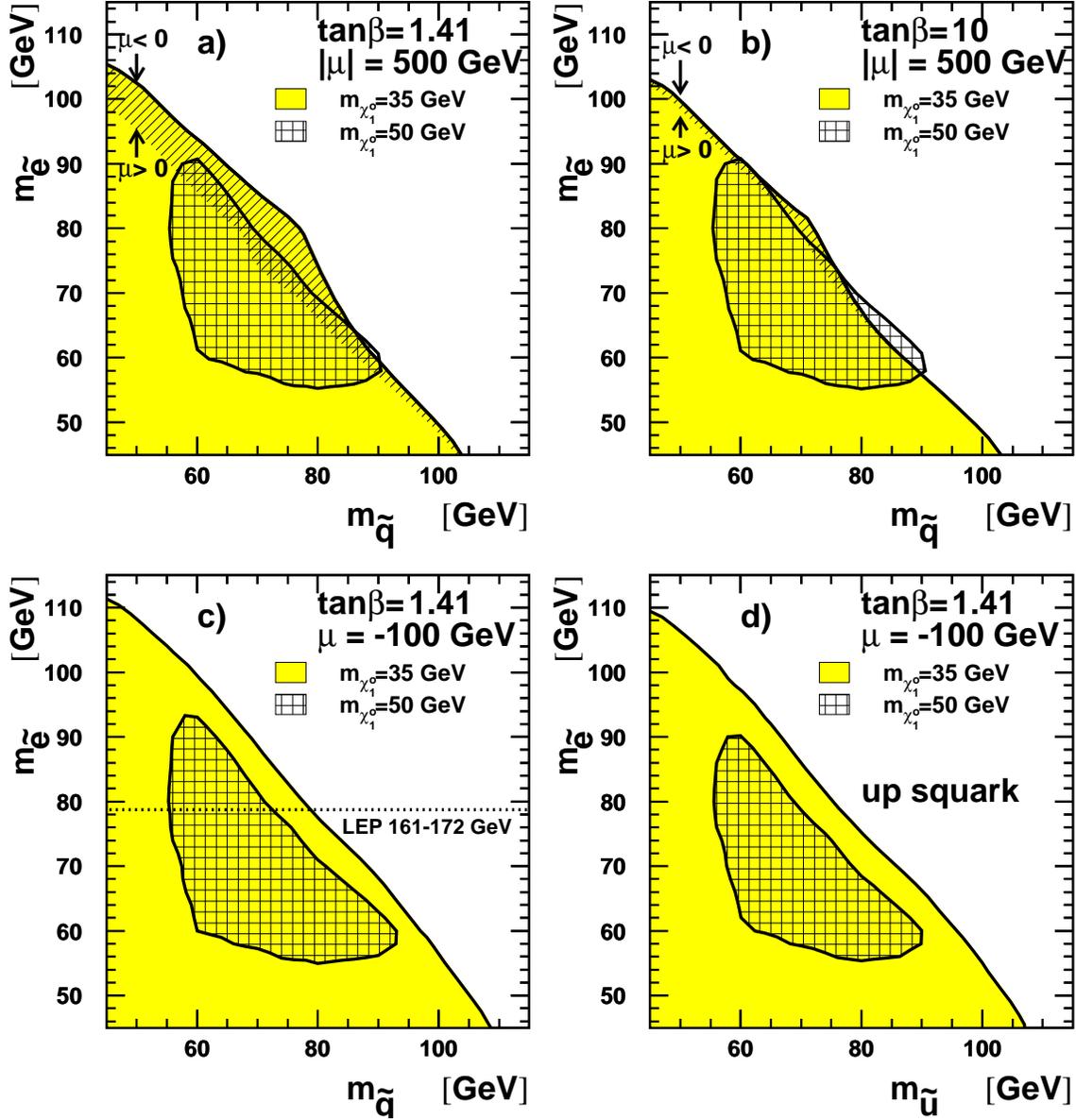,height=17cm}
                    }
                                 }
                                 
         \caption{\label{f3} Excluded regions at the 95\% CL
                  in the plane defined by the selectron and squark mass,
                  for $m_{\chi_{1}^{0}}=35$ GeV (grey area) 
                  and $m_{\chi_{1}^{0}}=50$ GeV (double-hatched area).
                  In a) ($\tan \beta =1.41$) and b) ($\tan \beta =10$) the 
                  limits are for $|\mu| = 500$ GeV. For $\mu<0$ the 
                  excluded region includes also the single-hatched area. 
                  The limits obtained for $\mu=-100$ GeV and 
                  $\tan \beta = 1.41$ are shown in c), where LEP limits 
                  on $m_{\sel}$ are also given.
                  The limits for the up squark alone, for the same values of
                  $\mu$ and $\tan \beta$, are shown in d).                   
}
\end{figure}

Figure~\ref{f3} shows the excluded regions in the plane $m_{\tilde e}$,
$m_{\tilde q}$ for fixed values of $m_{\chi^0_1}=35, 50$ GeV and 
for different combinations of $\mu$ and $\tan\beta$.
The excluded area is approximately triangular, defined by a line of constant 
$m_{\sel}+m_{\squ}$ and by two lines parallel
to the coordinate axes due to the low efficiency for $\Delta m < 10$ GeV. 
In figures~\ref{f3}~a) and b) the limits are given for the large $\mu$ region 
($|\mu|=500\,\mbox{GeV}$) and $\tan \beta =$1.41, 10. 
The limits obtained at $\tan \beta =1.41$ and $\mu=-100$ GeV
are shown in figure~\ref{f3} c).

In the range $45<({m_{\sel}+m_{\squ})/2}<85\,{\rm GeV}$ 
the up quark contribution to the cross section ranges between 70\% and 90\%
because it dominates the parton densities at high-$x$. 
The limits on the $\tilde u$ mass, assuming all the other squarks to be
much heavier, are only $\sim 2$ GeV below the limit obtained
for degenerate squark masses. This is shown as an example in
figure~\ref{f3}~d), to be compared with figure~\ref{f3} c).

\section{Comparison with other experiments}
\label{sec:comparison}

A search for the production of selectrons and squarks at HERA has been 
published by H1 \cite{h1}. The present analysis substantially improves 
these limits due to the seven-fold increase in the integrated luminosity. 

Published limits from LEP at $\sqrt{s}=161-172$ GeV exclude 
selectrons with masses lower than $\sim 80$ GeV \cite{lep}.
The present analysis goes beyond this limit for squark masses below 
$\sim 80$ GeV.
LEP experiments have also reported limits on the stop and on the sbottom 
squarks \cite{stop} of $m_{\tilde t},m_{\tilde b}\stackrel{>}{\sim} 75$ GeV.
The limits given in our analysis are (for $m_{\sel}\simeq m_{\squ}$)
at the same level as those obtained by combining the LEP limits on $m_{\sel}$ 
with those on $m_{\tilde t}$ and $m_{\tilde b}$ extended to the other squarks,
including $\tilde u$.
Present limits on the mass of the lightest neutralino \cite{neu}
exclude $m_{\lsp}<25$ GeV for $\tan\beta \sim 1$ and $m_{\lsp}<40$ GeV for 
large $\tan\beta$.

Strong limits on the squark mass have been obtained from the $p\bar{p}$ 
experiments at the Tevatron \cite{tevatron}
and are complementary to those obtained in the present analysis.
The Tevatron experiments adopt strong cuts
on $\PT$ and select final states with many high-$P_{\rm t}$ jets or leptons
to exclude the QCD background. Therefore, they are more 
sensitive to squark cascade decays than to the direct decay 
$\squ \rightarrow q \lsp$, especially
when the mass difference $m_{\squ} - m_{\lsp}$ is small ($\sim 20$ GeV). 
On the other hand, the search described in this paper looks only for direct 
decays and, due to the relatively low cuts on $\PT$ (14 GeV)
and $P_t^h$ (4 GeV), is sensitive to smaller mass differences.
Limits on SUSY are often expressed in terms of the
minimal supergravity model (mSU\-GRA) in which the masses of scalars 
and fermions are no longer independent parameters \cite{sugra}.
In this model the squarks are always heavier than selectrons
and gluinos, and the dominant squark decay is to gluino and quark.
In the region where the squark is lighter than the gluino, 
the Tevatron experiments give limits
under more general assumptions than mSUGRA but keep the GUT inspired 
assumption that links the masses of the neutralinos and the charginos to that
of the gluinos. In this framework limits cannot be set on light squarks 
($\sim 80$ GeV) when the gluino is too heavy ($\sim 500$ GeV), due to the lack
of sensitivity at small $m_{\squ} - m_{\lsp}$. 

The present analysis investigates a different region of the MSSM 
parameter space,
where the gluinos are heavier than the squarks and the squarks masses are 
independent of the selectron masses, allowing also for 
$m_{\squ}\leq m_{\sel}$. Moreover, no relationships are assumed between 
$m_{\lsp}$ and $m_{\tilde g}$.

\section{Conclusions}
\label{sec:conclus}

We have searched for the SUSY process
$e^+p\rightarrow \sel^{+} \squ {\rm X}\,(\sel^{+}\rightarrow e^+\lsp \, ,
\squ \rightarrow q \lsp)$ using 46.6~pb$^{-1}$ of data at an
$e^{+}p$ center-of-mass energy of 300~GeV recorded with the ZEUS detector at 
HERA. One candidate event is found while 
$1.99_{-0.84}^{+0.57}$ are expected
from Standard Model processes.
The upper limit on the production cross section 
times branching ratios is found to be $\sigma\times B\,<\,0.5\,\mbox{pb}$ 
at the 95\% CL for mass differences $m_{\sel}-m_{\lsp}$ and
$m_{\squ}-m_{\lsp}$ greater than 10 GeV.
Excluded regions in the MSSM parameter space have been derived.
We exclude $({m_{\sel}+m_{\squ})/2}\,<77\,\mbox{GeV}$ at the 95\% CL for 
$m_{\lsp}\,=\,40\,\mbox{GeV}$ and large values of the MSSM parameter
$|\mu|$. 
The process is dominated by the $\tilde u$ contribution and the 
exclusion limit is 75 GeV when only the $\tilde u$ squark is considered.

\section*{Acknowledgements}
We appreciate the contributions to the construction and maintenance of
the ZEUS detector by many people who are not listed as authors.  We
especially thank
 the DESY computing staff for providing the data analysis
environment and
 the HERA machine group for their
outstanding operation of the collider. Finally, we thank the DESY
directorate for strong support and encouragement.

\vfill\eject

\begin{table}
\begin{center}
\begin{tabular}{|l||c|l|c|l||l|} \hline
Process     & Expected Events  & Stat.  &  Theor. & Syst. & Total \\   \hline \hline
NC DIS      &  0.60   &  $\pm 0.29$     & $\pm 0.03$   & $_{-0.67}^{+0.06}$
                                                    & $_{-0.73}^{+0.30}$    \\ \hline
CC DIS      & 0.18   & $\pm 0.07$       & $\pm 0.01$ & $_{-0.10}^{+0.24}$
                                                    & $_{-0.12}^{+0.25}$   \\ \hline

W           & 1.04   &  $\pm 0.04$      & $\pm 0.36$ & $_{-0.12}^{+0.09}$
                                                    & $_{-0.38}^{+0.37}$  \\ \hline
$l^+l^-$    & 0.17    & $\pm 0.12$      & $\pm 0.02$ & $_{-0.08}^{+0.17}$
                                                    & $_{-0.14}^{+0.21}$\\ \hline 
 Total      & 1.99    & $\pm 0.32$      & $\pm 0.36$ & $_{-0.69}^{+0.31}$
                                                    & $_{-0.84}^{+0.57}$  \\ \hline
\end{tabular}
\end{center}
\caption{\label{t1}Background expectation from MC calculations. 
The table shows the background process, 
the expected number of events normalized to data
luminosity, the statistical error related to the MC statistics,
the error due to the theoretical uncertainty on the cross section,
the systematic error and the total error.}

\end{table}
 
\end{document}